\documentclass[longbibliography,12pt]{iopart}
\usepackage{multirow}
\usepackage{graphicx}
\usepackage{epstopdf}
\usepackage{bm}
\usepackage{color}
\usepackage{CJKutf8}
\definecolor{nblue}{rgb}{0.0, 0.0, 1.0}
\definecolor{magenta}{rgb}{0.79, 0.08, 0.48}

\begin{document}
\title{Superconductivity in CeBeH$_{8}$ and CeBH$_{8}$ at moderate pressures }

\author{Yu Hou$^1$, Bin Li$^{2}$, Yan Bai$^{1}$, Xiaofeng Hao$^1$, Yeqian Yang$^1$, Fengfeng Chi$^2$, Shengli Liu$^2$, Jie Cheng$^{2}$, Zhixiang Shi$^3$}
\address{$^1$ College of Electronic and Optical Engineering, Nanjing University of Posts and Telecommunications, Nanjing 210023, China}
\address{$^2$ New Energy Technology Engineering Laboratory of Jiangsu Province and School of Science, Nanjing University of Posts and Telecommunications, Nanjing 210023, China}
\address{$^3$ School of Physics, Southeast University, Nanjing 211189, China}
\ead{libin@njupt.edu.cn (B Li) and {liusl@njupt.edu.cn} (S Liu)}

\begin{abstract}
  High-pressure structural searches of superhydrides CeBeH$_8$ and CeBH$_8$ were performed under ambient pressure up to 300 GPa. We identify $Fm\overline{3}m$-CeBeH$_8$ with a superconducting transition temperature $T_{c}$ of 56 K at 10 GPa. Two more phases with spacegroup $R\overline{3}m$ and $C2/m$, were investigated within the increasing pressures. CeBH$_8$ shows a similar phase transition process as CeBeH$_8$ but with higher transition pressures and higher $T_c$. $Fm\overline{3}m$-CeBH$_8$ is predicted to be superconducting above 120 GPa with a maximum $T_{c}$ of 118 K at 150 GPa. $R\overline{3}m$-CeBH$_8$ and $C2/m$-CeBH$_8$ are dynamically stable above 120 GPa and 100 GPa, respectively. The maximum $T_{c}$ is 123 K at 195 GPa for $R\overline{3}m$-CeBH$_8$, and 115 K at 350 GPa for $C2/m$-CeBH$_8$. Our work enriches the family of ternary hydrides and may provide a useful guideline for further search for superconducting hydrides at low and moderate pressures.
\end{abstract}
\noindent{\bf Keywords:}
\noindent{\it Superconductivity, Hydrides, High pressures\/}\\
\submitto{JPCM}
\maketitle

\section{Introduction}
The pursuit of room-temperature superconductivity has been a major challenge since Onnes discovered the phenomenon of superconductivity in 1911. The Bardeen-Cooper-Schrieffer (BCS) theory\cite{1957bcs} well explains the mechanism of traditional superconductors and provides a route for discovering and predicting high $T_{c}$ materials. According to the BCS theory, the superconducting transition temperature is directly proportional to its Debye temperature, which is inversely proportional to the mass of the element. It can be predicted that the lightest element hydrogen may have an ultra-high superconducting transition temperature. On the other hand, the hydride systems are easier to be metalized than the pure hydrogen, due to the presence of non-hydrogen elements can exert chemical pressure on the sublattice of hydrogen\cite{N2004Hydrogen}. Metallization and superconductivity of hydrogen-rich materials under high pressures is an effective way to realize metal hydrogen and high-$T_{c}$ superconductor (HTS).

The most initiated studies have mainly focused on binary hydrides, such as H$_3$S (maximum $T_{c}$ $\sim$ 203 K)\cite{A2015Conventional}, LaH$_{10}$ (maximum $T_{c}$ $\sim$ 260 K)\cite{somayazulu2019evidence,drozdov2019superconductivity}, and cerium polyhydride CeH$_{10}$ and CeH$_{9}$\cite{ceh2019,ceh2021} et al.. After trying a large number of binary hydrides, the gravity of the study was gradually shifted to the ternary hydrides. Compared to the binary hydrides, exploring the high $T_{c}$ superconducting in the ternary hydrides greatly expanded the phase spaces and provided new chemical and physical avenues, which in principle could improve the superconducting transition temperature and stability. The additional element degree of freedom would lead to new predicted high $T_{c}$ superconductors in compressed ternary hydride systems, such as CaYH$_{12}$ ($T_{c}$ = 258 K at 200 GPa)\cite{liang2019potential}, LiPH$_6$ ($T_{c}$ = 150-167 K at 200 GPa)\cite{mu2019uncovering}, and Li$_2$MgH$_{16}$ ($T_{c}$ = 473 K at 250 GPa)\cite{sun2019route}. A carbonaceous sulfur hydride was experimentally verified with $T_{c}$ around 280 K at 272 GPa\cite{snider2020room}, which is the highest superconducting transition temperature that has been experimentally reported for hydrogen compound superconducting materials to date.

Although great results have been achieved in the search for HTS hydrides, the tremendous pressure required to stabilize the superconducting phase hinders any practical applications. The next challenge in materials research is to find materials with superconducting at or near ambient pressure. Recently, LaBH$_8$ was reported to achieve superconducting at a pressure of 50 GPa with a transition temperature of 126 K\cite{di2021bh} and a pressure of 20 GPa with a transition temperature of 185 K\cite{2022Design}. Its exceptional superconducting properties derive from a metallic hydrogen lattice, stabilized by the mechanical pressure exerted by an inert boron and lanthanum scaffold, with a mechanism analogous to binary sodalite-clathrate hydrides.

In this work, we show an approach that bring the stability pressure of HTS hydrides closer to ambient pressure by combining different elements to enhance the chemical precompression observed in binary hydrides. Through a systematic search of the phase diagram of Cerium-Beryllium-Hydrogen(Ce-Be-H) and Cerium-Boron-Hydrogen(Ce-B-H) system from ambient pressure up to 300 GPa, we finally determine three superconducting phases: $Fm\overline{3}m$, $R\overline{3}m$ and $C2/m$ for both CeBeH$_8$ and CeBH$_8$. The electronic band structures, phonon spectra and electron-phonon (\emph{e-ph}) coupling are studies. Higher symmetry usually beneficial to a lower transition pressure and better superconducting, and the middle element X in CeXH$_8$ affect the stable pressure range and the corresponding $T_c$. Boron hydrides have higher superconducting temperatures but higher stable pressures, while beryllium hydrides can survive at low pressures but the superconducting temperatures are suppressed.

\section{Methods}
We used a machine-learning-based crystal structure predictor package CRYSTREE\cite{crystree1,crystree2} to predict the stable structures of Ce-Be-H and Ce-B-H systems at pressures of 0, 50, 100, 150, 200, 250 and 300 GPa. The results are verified by using the evolutionary crystal structure prediction method USPEX\cite{2006USPEX,ux3,ux1}. The electronic structure calculations with high accuracy for the stable structures were performed using the full-potential linearized augmented plane wave (FP-LAPW) method implemented in the WIEN2K code~\cite{Wien2k}. The generalized gradient approximation (GGA)\cite{GGA} was applied to the exchange-correlation potential calculation. The crystal structures were visualized by VESTA\cite{vesta}. Fermi surfaces are visualized using XCRYSDEN\cite{xcrysden} and Femisurfer\cite{fermisurfer}. The phonon calculations including \emph{e-ph} couplings were carried out by using a density functional perturbation theory (DFPT)\cite{gonze1997dynamical} approach through the Quantum-ESPRESSO\cite{giannozzi2009quantum} program. The pseudopotentials are selected from standard solid state pseudopotentials\cite{sssp} library with Perdew-Burke-Ernzerhof(PBE) functional. The cutoffs were chosen as 60 Ry for the wave functions and 600 Ry for the charge density. The electronic integration was performed over an 8$\times$8$\times$8 k-point mesh. Dynamical matrices and the electron-phonon couplings were calculated on a 4$\times$4$\times$4 q-point grid. A dense 24$\times$24$\times$24 grid was used for evaluating an accurate electron-phonon interaction matrix. Finally, $T_{c}$ is obtained by the modified McMillan equation\cite{mcmillan1968transition}.

\section{Results and discussion }

\begin{figure*}
\begin{center}
\includegraphics[width=16cm]{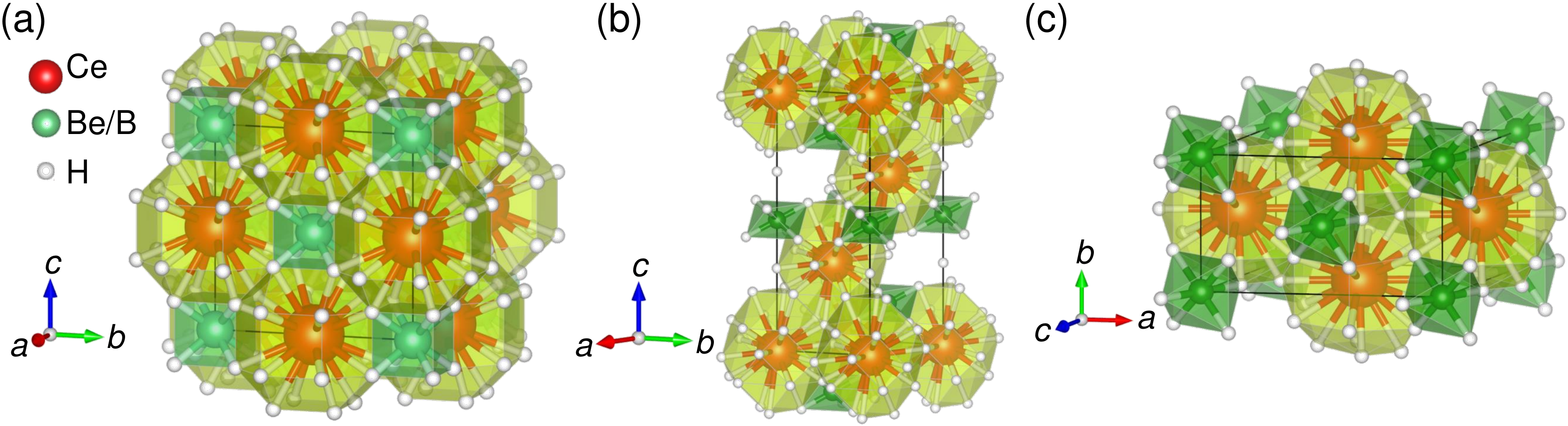}
\caption{The crystal structures of (a) $Fm\overline{3}m$, (b) $R\overline{3}m$ and (c) $C2/m$ phases, Red spheres represent cerium atoms, green represents beryllium or boron atoms, and white spheres represent hydrogen atoms.}
\label{struct}
\end{center}
    \end{figure*}

The crystal structures of $Fm\overline{3}m$, $R\overline{3}m$, and $C2/m$-CeXH$_8$ (X=B, Be) are shown in Fig. \ref{struct}. Red spheres represent cerium atoms, green ones represent beryllium or boron atoms, and white spheres represent hydrogen atoms. For the $Fm\overline{3}m$ phase, Ce and Be/B atoms occupy 4$b$ and 4$a$ Wyckoff positions, respectively, while H atoms locate at the 32$f$ position. The hydrogen atoms construct spherical polyhedrons with Ce atom located at the center, and a series of small hydrogen cubes with centered Be/B atoms, are inserted in the polyhedrons. The structure of $Fm\overline{3}m$-CeXH$_8$ is similar to sodalite-like hydride, e.g., CeH$_{10}$, in which guest atoms Ce act as a scaffold to exert mechanical pressure on the hydrogen sublattice\cite{2012Superconductive,2017Hydrogen}. In CeXH$_8$, Be/B atoms fill the interstitial sites between the second-nearest Ce atoms, efficiently filling the remaining voids in the CeH$_{10}$ structure\cite{ceh2019,ceh2021}. In the $R\overline{3}m$ structure, Ce and Be/B atoms occupy 3$a$ and 3$b$ positions, respectively, while H located at positions 18$h$ and 6$c$. In $C2/m$-CeXH$_8$, Ce atom occupies the 2$d$ position, Be/B occupies the 2$a$ position, and H atoms locate at 8$j$ and 4$i$ positions. {The enthalpy curves of the different phases can be found in the Supplementary material. When pressure less than 60 GPa for CeBeH$_8$, and 200 GPa for CeBH$_8$, $C2/m$ and $R\overline{3}m$ phases exhibit lower enthalpies, then these three phases coexist with similar enthalpies up to the maximum pressure we studied. }

\begin{figure*}
\includegraphics[width=16cm]{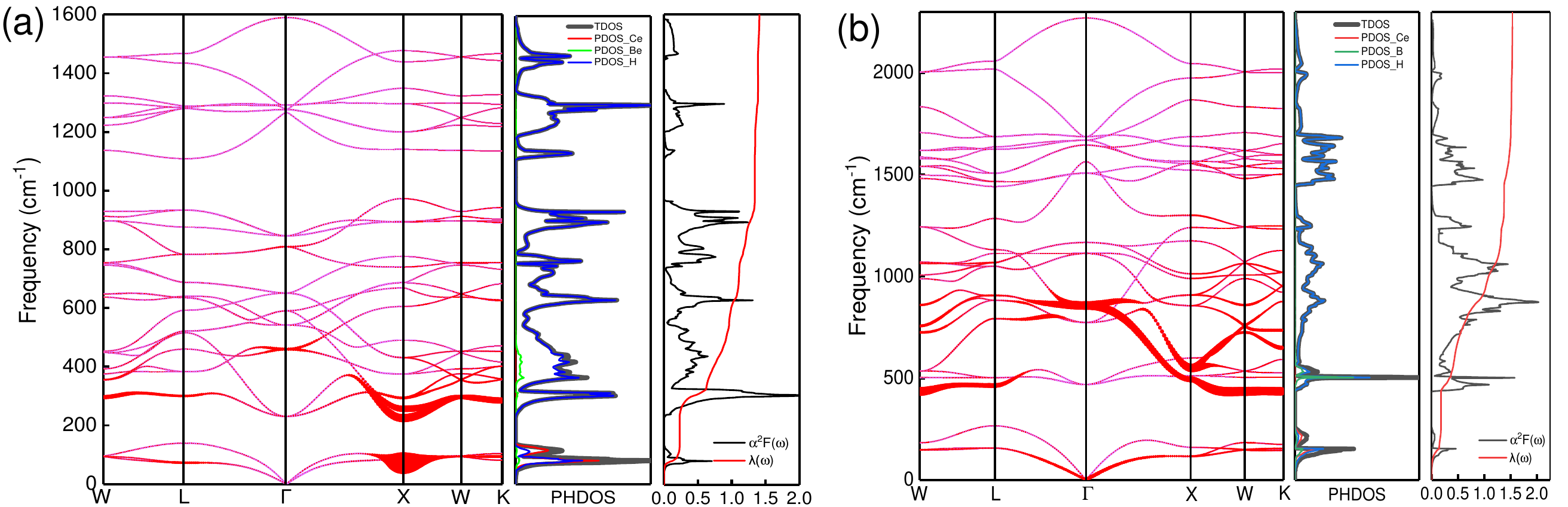}
\caption{Phonon dispersion relations, projected phonon densities of states (PHDOS), and Eliashberg spectral function for (a)  $Fm\overline{3}m$-CeBeH$_8$ at 10 GPa and (b)  $Fm\overline{3}m$-CeBH$_8$ at 150 GPa. Phonon dispersion is represented by purple thin lines,  decorated with the \emph{e-ph} coupling $\lambda_{\nu\textbf{q}}$ {in red circles}.} \label{phonon}
\end{figure*}

 We show the phonon dispersions and projected phonon densities of states (PHDOS) for $Fm\overline{3}m$-CeBeH$_8$ at 10 GPa and $Fm\overline{3}m$-CeBH$_8$ at 150 GPa in Fig. \ref{phonon}. The point group at $\Gamma$ point is $O_{h}(m\bar{3}m)$, the first three modes are triple-degenerate acoustic modes, and the rest are optical modes. The irreducible representation is $\Gamma$ = $4T_{1u}\oplus T_{2u}\oplus T_{1g}\oplus 2T_{2g}\oplus E_u\oplus E_g\oplus A_{1g}\oplus A_{2u}$. The phonon dispersions are decorated by the \emph{e-ph} coupling coefficients $\lambda_{\nu\textbf{q}}$ for the phonon branch $\nu$ at $\textbf{q}$ point. The phonon spectra for $R\overline{3}m$-CeBeH$_8$ and $C2/m$-CeBeH$_8$ can be found from Fig. S1 and Fig. S2 in the supplemental materials. From the $\lambda_{\nu\textbf{q}}$ distribution (red circles) in Fig. \ref{phonon}(a), it can be seen that strong \emph{e-ph} couplings are mostly distributed on the acoustic modes and low-frequency optical modes near $X$ point. The PHDOS curves in the middle panel demonstrate that the low-frequency part is mainly contributed by the vibrations of Ce atoms. For CeBH$_8$ in Fig. \ref{phonon}(b), strong \emph{e-ph} couplings mainly distribute on optical branches in the middle- and low-frequency regions from 500 to 1000 cm$^{-1}$. The low-frequency acoustic branches are mainly contributed by the vibrations of Ce and B atoms. A significant phonon peak is located at 500 cm$^{-1}$, which is contributed by the vibrations of B atoms. We also show the Eliashberg function $\alpha^{2}F(\omega)$ and the integration of \emph{e-ph} coupling $\lambda({\omega})$ in the right panels of Fig. \ref{phonon} (a) and (b). Integrating the Eliashberg function we obtain the \emph{e-ph} coupling $\lambda=2\int \alpha^2F(\omega)\omega^{-1}d\omega$ and logarithmically averaged phonon frequency $\omega_{ln}=exp[2\lambda^{-1}\int d\omega\alpha^{2}F(\omega)\omega^{-1}log\omega]$. From the \emph{e-ph} coupling curves in the right panels of Fig. \ref{phonon} (a) and (b), $\lambda({\omega})$ under 1000 cm$^{-1}$ contribute more than 90\% of total \emph{e-ph} coupling for both $Fm\overline{3}m$-CeBeH$_8$ and $Fm\overline{3}m$-CeBH$_8$.

\begin{figure*}
\begin{center}
\includegraphics[width=16cm]{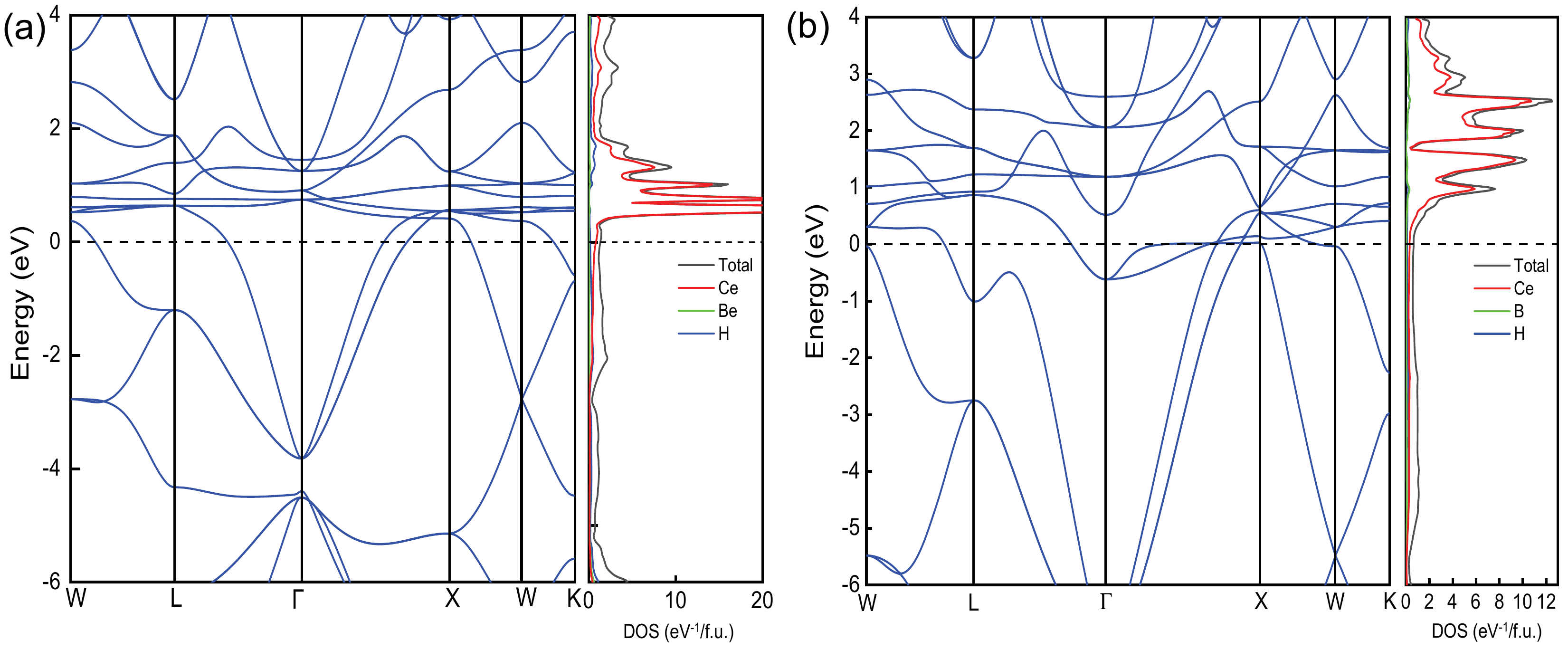}
\caption{Electronic band structure and atom-projected density of states for (a)  $Fm\overline{3}m$-CeBeH$_8$ at 10 GPa and (b)  $Fm\overline{3}m$-CeBH$_8$ at 150 GPa. Atom-projected density of states in units of eV$^{-1}$/f.u.. Projection onto Ce, Be/B, and H are shown in green, blue, and red, respectively. The zero of the energy scale is the Fermi level.} \label{bands}
\end{center}
\end{figure*}

In Fig. \ref{bands} we show the electronic band structures along with the atom-projected density of states (DOS) in units of eV$^{-1}$/formula unit (f.u.) for $Fm\overline{3}m$-CeBeH$_8$ at 10 GPa and $Fm\overline{3}m$-CeBH$_8$ at 150 GPa. It demonstrates that all of the predicted structures exhibit metallic behavior by evidence of bands crossing the Fermi level. Whether in Ce-Be-H or Ce-B-H systems, the Ce atoms dominate the DOS around the Fermi level. The large van Hove peaks appearing above the Fermi level would provide a large density of states, which is beneficial to the superconductivity through the BCS theory. The electronic band structure and density of states for $R\overline{3}m$-CeBeH$_8$ can be found in Fig. S3 in the supplemental materials.

\begin{figure*}
\begin{center}
\includegraphics[width=14cm]{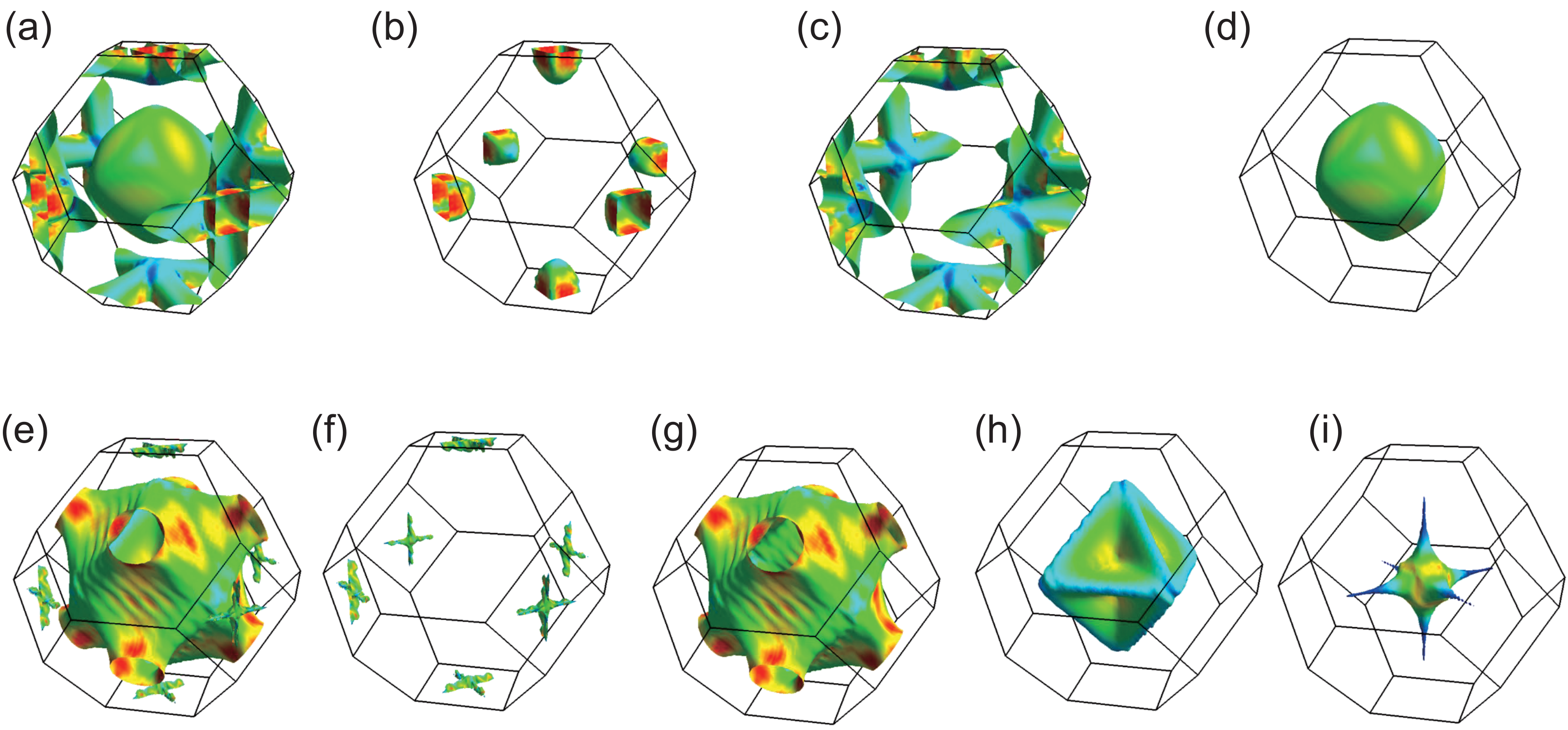}
\caption{(a) Fermi surfaces of $Fm\overline{3}m$-CeBeH$_8$ at 10 GPa, decorated with Fermi velocity. Three separate parts which construct the whole Fermi surface are shown in (b-d). (e) Fermi surfaces of $Fm\overline{3}m$-CeBH$_8$ at 150 GPa, (f-i) is the four separate parts.} \label{fermi}
\end{center}
\end{figure*}

Figure \ref{fermi} shows the Fermi surfaces for $Fm\overline{3}m$-CeBeH$_8$ and $Fm\overline{3}m$-CeBH$_8$, shadowed with the Fermi velocity distribution. The Fermi velocities increase with the color gradually changing from blue to red. The Fermi surface of CeBeH$_8$ at 10 GPa consists of three parts, a large electron-like sphere surrounding the $\Gamma$ point, six cross-shaped sheets and six small hole pockets around the Brillouin zone corners. The Fermi surface of CeBH$_8$ at 150 GPa consists of four parts. From figure \ref{fermi} (f-i), we can find a series of windmill-shaped sheets located near the $X$ points, an irregularly shaped body with eight circular cavities located around $\Gamma$, a regular octahedron with recessed inward surfaces, and a small sphere with six sharp spikes at $\Gamma$.

In order to estimate the superconducting $T_c$ of CeBeH$_8$ and CeBH$_8$ under pressures, we performed the linear response calculations on \emph{e-ph} properties, and calculated the critical temperature through the Allen-Dynes modified {McMillan} formula\cite{mcmillan1968transition,Allen1975}:

\begin{equation}
T_{c}=\frac{\omega_{ln}}{1.2}\textrm{exp}\left[-\frac{1.04(1+\lambda)}{\lambda-\mu^*(1+0.62\lambda)}\right],
\end{equation}

\noindent where $\omega_{ln}$ is the logarithmically averaged phonon frequency, $\lambda$ is the \emph{e-ph} coupling constant, and $\mu^*$ is the Coulomb pseudopotential which is set to be 0.1 in the calculations. Detailed calculation results are shown in {Figure \ref{phases} and Table S1 in Supplementary materials}. {The superconducting $T_c$s of various structural phases for CeBeH$_8$ are remarkably consistent over the whole pressure range. $T_c$ for each phase of CeBeH$_8$ gradually decreases with the increase of pressure. However, the $T_c$ variation in CeBH$_8$ is rather noticeable. The $T_c$ curve of $R\overline{3}m$-CeBH$_8$ is close to the $C2/m$ curve at the beginning, but suddenly rises after 200 GPa and coincides with the curve of $Fm\overline{3}m$-CeBH$_8$ at 300 GPa. It is notable that the Be-base and B-base systems share the exact same structural phases but emerge with different tendencies of superconducting temperatures at high pressures. The Be element is crucial in reducing the stable pressure, as shown by a previous work on LaBeH$_8$ that found it to be dynamically stable down to 20 GPa\cite{2022Design}. The calculation results shown in Table S1 of the Supplemental material indicate that Be-based compounds generally have lower \emph{e-ph} coupling and $\omega_{ln}$ than B-based compounds, which is the main reason for their lower superconducting critical temperatures.}

$Fm\overline{3}m$-CeBeH$_8$ can achieve dynamic stability above 10 GPa, with a maximum $T_{c}$ of 56 K. $T_{c}$ decreases gradually with the increasing pressures. $R\overline{3}m$-CeBeH$_8$ is stable above 20 GPa and has a maximum $T_{c}$ of 30 K at 20 GPa. The maximum $T_{c}$ of $C2/m$-CeBeH$_8$ is 12 K at 50 GPa and stable above 50 GPa. For CeBeH$_8$, the increasing pressures suppress the structural symmetry and the superconductivity, the phases with higher stable pressures usually have lower superconducting transition temperatures. $Fm\overline{3}m$-CeBH$_8$ reaches the maximum $T_{c}$ of 118 K at 150 GPa, $R\overline{3}m$-CeBH$_8$ and $C2/m$-CeBH$_8$ reaches maximum $T_{c}$ of 123 K at 195 GPa and 115 K at 350 GPa, respectively.

\begin{figure}
\begin{center}
\includegraphics[width=14cm]{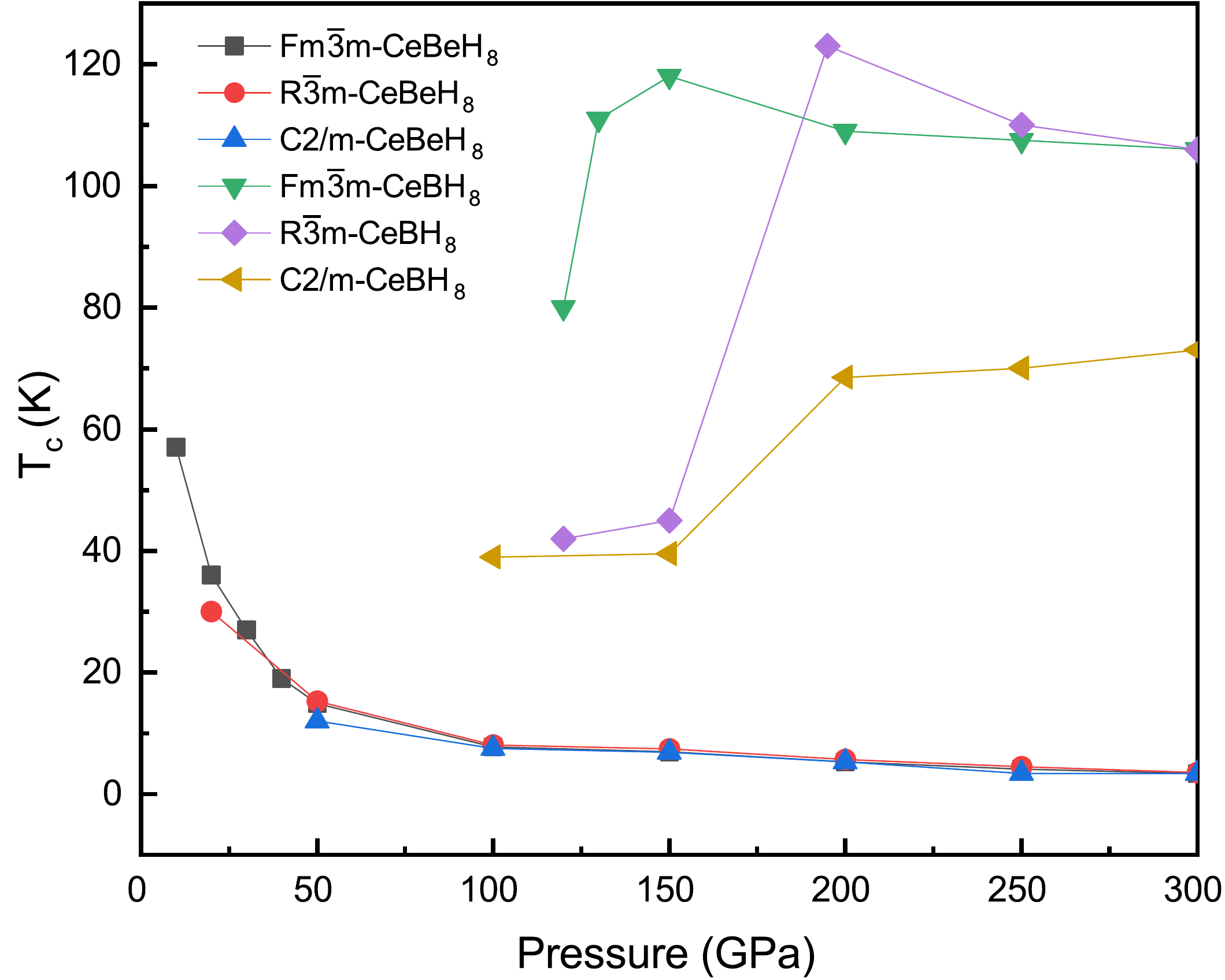}
\caption{Variation of $T_c$ with pressure in different phases of Ce-Be-H and Ce-B-H systems.}
\label{phases}
\end{center}
\end{figure}

\section{Conclusion}

In summary, using crystal structure prediction and first-principles calculation, we have found two new cerium hydride superconductors, CeBeH$_8$ and CeBH$_8$ with $Fm\overline{3}m$, $R\overline{3}m$ and $C2/m$ symmetries. The  CeXH$_{8}$ structures can be adjusted by screening the X elements, so as to achieve more effective stacking and increase the chemical pre-compression effect. The exceptional superconducting properties derive from a metallic hydrogen lattice, stabilized by the mechanical pressure exerted by beryllium/boron and cerium scaffold, with a mechanism analogous to binary sodalite-clathrate hydrides. Through the selection and adjustment of the X element, the stable pressure for $Fm\overline{3}m$-CeBeH$_8$ is reduced to 10 GPa, which is an unprecedented low pressure in ternary hydrides. Our results confirm the accuracy of the machine-learning-based crystal structure prediction and represent a direction in the field of superconductivity in the future. The search for hydride superconductors that can be stable at low pressure, even at ambient pressure, may have the possibility of realizing the practical application.

\ack{
This work is supported by the National Key R\&D Program of China (Grant No. 2018YFA0704300), the National Natural Science Foundation of China (Grants No. U1932217, 12175107), and NUPTSF (Grant No. NY219087, NY220038). Some of the calculations were performed on the supercomputer in the Big Data Computing Center (BDCC) of Southeast University.\\}


\end{document}